\documentclass{aa}

\usepackage{schemata}
\usepackage{xspace}
\usepackage{graphicx}
\usepackage[table,dvipsnames]{xcolor}
\usepackage{txfonts}
\usepackage[pdfencoding=auto,psdextra]{hyperref}
\usepackage{placeins}
\usepackage{orcidlink}
\usepackage{color}
\usepackage{soul}
\usepackage{ulem}[normalem]

\newcommand{\hstfull}{\textit{Hubble Space Telescope}\xspace}
\newcommand{\hst}{\textit{HST}\xspace}

\newcommand{\gaia}{\textit{Gaia}\xspace}
\newcommand{\hipparcos}{\textit{Hipparcos}\xspace}
\newcommand{\qfit}{\texttt{QFIT}\xspace}

\newcommand{\maspixel}{mas pixel$^{-1}$\xspace}
\newcommand{\masyr}{mas yr$^{-1}$\xspace}

\begin{document}

\title{Astrometric Reconnaissance of Exoplanetary Systems (ARES)}
\subtitle{I. Methodology validation with \hst point-source images of Proxima\,Centauri}
\titlerunning{ARES. I}

\authorrunning{Libralato et al.}

\author{M.~Libralato\thanks{\email{mattia.libralato@inaf.it}}\inst{\ref{aff1}}\orcidlink{0000-0001-9673-7397}
\and L.~R.~Bedin\inst{\ref{aff1}}\orcidlink{0000-0003-4080-6466}
\and A.~Burgasser\inst{\ref{aff2}}\orcidlink{0000-0002-6523-9536}}

\institute{INAF - Osservatorio Astronomico di Padova, Vicolo dell'Osservatorio 5, Padova, I-35122, Italy\label{aff1}
\and
Department of Astronomy \& Astrophysics, University of California, San Diego, La Jolla, California 92093, USA\label{aff2}
}

\date{Received 17 October 2025 / Accepted 28 November 2025}

\abstract{
We present the first results of the \textit{Astrometric Reconnaissance of Exoplanetary Systems} (ARES) project, aimed at validating and characterizing candidate exoplanets around the nearest systems using multi-epoch \hstfull (\hst) data. In this first paper, we focus on Proxima\,Centauri, leveraging archival and recent \hst observations in point-source imaging mode. We refine the geometric-distortion calibration of the \hst detector used, and develop a robust methodology to derive high-precision astrometric parameters by combining \hst measurements with the \gaia DR3 catalog. We determine Proxima’s position, proper motion, and parallax with uncertainties at the $\sim$0.4-mas, 50-$\mu$as yr$^{-1}$, and 0.2-mas level, respectively, achieving consistent results with what measured by \gaia within $\sim$$1\sigma$. We further investigate the presence of the candidate exoplanet Proxima\,c by analyzing the proper-motion anomaly derived from combining long-term \hst-based and short-term \gaia astrometry. Under the assumption of a circular, face-on orbit, we obtain an estimated mass of $m_c = 3.4^{+5.2}_{-3.4}$ $M_\oplus$, broadly consistent with radial-velocity constraints but limited by our current uncertainties. These results establish the foundation for the next phase of ARES, which will exploit \hst spatial-scanning observations to achieve astrometric precisions of a few tens of $\mu$as and enable a direct search for astrometric signatures of low-mass companions.
}

\keywords{astrometry - stars: low mass - planets and satellites: detection}

\maketitle

\nolinenumbers

\section{Introduction}\label{sec:intro}

Astrometry is a niche technique in the field of exoplanet detection, but is potentially one of the most informative. The basis of the astrometric method is simple: the mutual gravitational interaction between a host star and its companion(s) makes them orbit around a common center of mass. For the host star, this orbital motion manifests as an astrometric ``wobble'' that deviates from the star’s normal proper and parallactic motions. Detection of an exoplanet by the host star’s astrometric wobble is similar in nature to the line-of-sight (LOS) radial-velocity (RV) method; however, because astrometric wobble is measured in two spatial dimensions instead of one, it does not suffer from the mass-inclination degeneracy of the RV method. Also, the astrometric method is not affected by other limitations of spectroscopic measurements like spectral-line asymmetries or gravitational redshift. By measuring the deviations from the expected motion of the host star on the sky, one can directly derive the perturber’s mass (if the host mass is known) and its orbital parameters. Alas, the astrometric wobble induced on a host star is proportional to the planet/host star mass ratio and is hence typically very small; it is also inversely proportional to the distance of the host star to the Earth, and is thus only relevant for the closest stars to the Sun.\looseness=-4

The \gaia Data Release 3 (DR3) release \citet{2016GaiaCit,2023A&A...674A...1G} has provided astrometry for more than a billion of sources, and the \gaia consortium has run a preliminary search for companions around nearby stars in the unpublished \gaia-DR3 astrometric time series \citep{2023A&A...674A..10H}. However, the least-massive objects (brown dwarfs and exoplanets) will be detectable by the astrometric method only around the brightest host stars. Low-mass companions to low-mass stars, the most common type of stars in the Milky Way, are generally too faint for \gaia measurements \citep{2014ApJ...797...14P}, as are stars in crowded fields or with extreme properties. The release of all the \gaia individual measurements with the DR4 (ca. December 2026) will allow the community to develop new tools to search for planets at the faint end of \gaia’s sensitivity. However, these detections will be difficult to validate as few other instruments can provide a comparable astrometric precision. Alternative methods and facilities are needed to both fill the low-mass gap left by \gaia as well as independently validate the discoveries \gaia will make. The ``Astrometric Reconnaissance of Exoplanetary Systems'' (ARES) project is designed to address this need through a challenging use of \hstfull (\hst) data. ARES is aimed at validating and characterizing candidate exoplanets orbiting two of the closest systems to the Sun: Proxima\,Centauri (Proxima) and Luhman 16AB (Luh 16AB).\looseness=-4

\begin{figure*}[t]
    \sidecaption
    \centering
    \includegraphics[width=12.9cm]{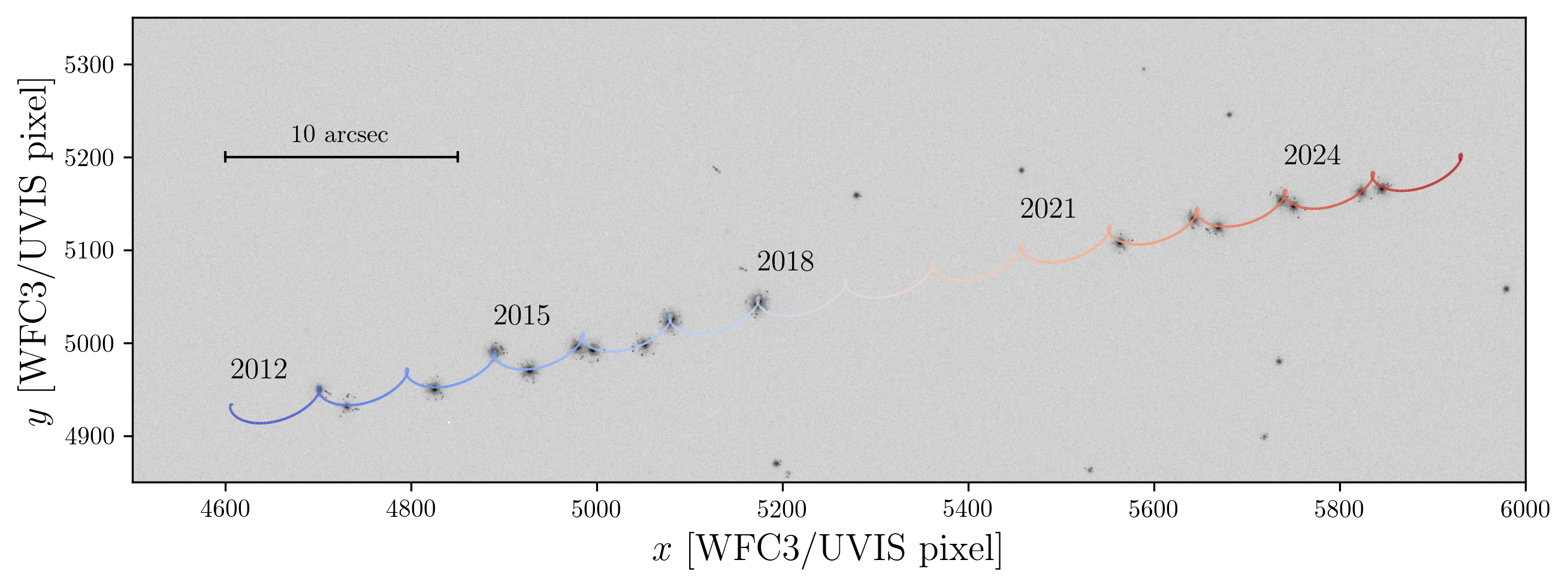}
    \caption{Composite image of Proxima obtained by co-adding all \hst exposures, properly normalized, from October 2012 to February 2025. The colored line represents the motion of Proxima predicted by our astrometric fit (see Sect.~\ref{sec:astromodel}), with the color changing from blue to white to red as the epoch increases (see also labels in the plot). North is up and East is to the left.\looseness=-4}
    \label{fig:fov}
\end{figure*}

\begin{table*}[t]
\vspace{0.5cm}
\centering
\caption{Log of the observations used in this work.}
\begin{tabular}{ccccc}
\hline
Observing Program & PI & Filter & $N_{\rm exp} \times t_{\rm exp}$ & Epoch \\
\hline
GO-12985 & Sahu  & F555W & $ 4 \times 1$\,s & October 2012 --    March 2013 \\
GO-13466 & Sahu  & F555W & $10 \times 1$\,s & October 2013 --    April 2014 \\
GO-13847 & Sahu  & F555W & $43 \times 1$\,s &   April 2015 --  January 2018 \\
GO-16653 & Bedin & F467M & $ 6 \times 5$\,s &  August 2022 --    March 2023 \\
GO-16983 & Bedin & F467M & $ 4 \times 5$\,s & October 2023 -- February 2024 \\
GO-17399 & Bedin & F467M & $ 4 \times 5$\,s &    July 2024 -- February 2025 \\
\hline
\end{tabular}
\tablefoot{F555W data were taken in subarray mode, with only chip 1 used. F467M observations were taken using the full array, and Proxima was imaged either on chip 1 or 2, depending on the detector's position angle.}
\label{tab:log}
\end{table*}

Proxima is a late-type M dwarf and the closest star to the Sun at $\sim$1.3 pc. It is known to host multiple exoplanets, all found by the LOS RV method. Proxima b \citep[discovered by][]{2016Natur.536..437A} has a minimum mass of $1.0 \pm 0.1$$M_\oplus$, a semi-major axis of $\sim$0.05 AU, and a period of $\sim$11.2 days \citep{2020SciA....6.7467D}, placing it within Proxima’s habitable zone. Proxima d has a minimum mass of $0.26 \pm 0.05$ $M_\oplus$, a semi-major axis of 0.03 AU, and a period of 5.1 days \citep{2022A&A...658A.115F}. A third exoplanet, Proxima\,c, was reported by \citet{2020SciA....6.7467D}, with a minimum mass of $5.8 \pm 0.1$ $M_\oplus$, a semi-major axis of $\sim$1.58 AU, and a period of $\sim$1900 days. However, the existence of Proxima\,c remains uncertain, even after subsequent direct imaging and astrometric observations \citep{2020A&A...638A.120G,2019KervellaPMa,2020KervellaProxima,2022KervellaPMa,2020RNAAS...4...86B}. As discussed in \citet{2020SciA....6.7467D}, the existence of Proxima\,c would be a challenge to various scenarios about the formation of super-Earth exoplanets. Confirming the existence of Proxima\,c is thus extremely important for exoplanet formation theory.\looseness=-4

Luh 16AB is a brown-dwarf binary discovered by \citet{2013ApJ...767L...1L} with the Wide-field Infrared Survey Explorer (WISE). At only $\sim$2 pc from the Sun, Luh 16AB is the closest brown dwarf and brown-dwarf binary to the Sun. Luh 16A is an L8 dwarf with a mass of $35.4 \pm 0.2$ $M_{\rm Jup}$, while Luh 16B is a T1 dwarf with a mass $29.4 \pm 0.2$ $M_{\rm Jup}$ \citep[both masses from][]{2024AN....34530158B}. This system straddles the transition between the L dwarf and T dwarf spectral classes, a phase of brown dwarf evolution characterized by significant changes in their atmospheric properties, including the formation of CH4 gas and the depletion of condensate clouds \citep[e.g.,][]{2013ApJ...772..129B,2013ApJ...770..124K,2021ApJ...906...64A}. High-precision astrometric measurements of the system have revealed perturbations of their proper, parallactic and binary orbital motion that may be related to a third, unseen body. The combined efforts of \citet{2014A&A...561L...4B}, \citet{2018A&A...618A.111L} and \citet{2017BedinLuhman,2024AN....34530158B} have ruled out candidate companions down to Neptunian masses, limited by their astrometric precisions.\looseness=-4

The ARES project plans to measure astrometric perturbations in the motion of these systems by unseen companions using multi-epoch \hst observations taken with the Ultraviolet and VISible (UVIS) channel of the Wide Field Camera 3 (WFC3). The best techniques in point-source imaging with WFC3/UVIS can deliver a precision of $\sim$300-400 $\mu$as \citet{2009BelliniWFC3,2011BelliniWFC3}, which is not enough for a statistically-significant detection of the astrometric wobble induced by the candidate exoplanets. We will instead use images obtained using the so-called ``spatial scanning'' mode, an observing mode that slews the telescope during the observation and spreads the light along a trail. The many samplings obtained in a trail enable astrometric precisions of tens of $\mu$as in a single exposure, far beyond what is possible with \hst using the ``classical'' point-source imaging \citep[e.g.,][]{2014Riess}.\looseness=-4

This first paper of the series describes how we improved the methodology described in \citet{2017BedinLuhman,2024AN....34530158B} to set up the foundations to fit the astrometric parameters of our targets. At this aim, we choose to focus on point-source images of Proxima for which we have a clear understanding of positional precision and systematics. Proxima is bright and well-measured in the \gaia DR3 catalog, so we can compare the astrometric parameters derived with our \hst data with those from \gaia. Future work will focus on the analysis of the images in spatial-scanning mode, and search for astrometric signatures of companions. The paper is organized as follows: Sect.~\ref{sec:datared} introduces the data set used and its reduction using state-of-the-art techniques for precise and accurate astrometry; Sect.~\ref{sec:astrometry} provides an overview about the setup of the astrometric reference frame and the differences with past works on the same topic, as well as the fit of the \textit{absolute} astrometric parameters; Sect.~\ref{sec:pma} presents an attempt to estimate the mass of Proxima\,c combining our and \gaia's proper motions (PMs), and discusses the limits of the point-source imaging. Finally, conclusions and our next steps are described in Sect.~\ref{sec:conclusions}.\looseness=-4

\section{Data reduction}\label{sec:datared}

Proxima\,Centauri was imaged by \hst through several archival and our own proprietary General Observer (GO) programs (Table~\ref{tab:log}). For this first part of the project, we made use of WFC3/UVIS data taken between 2012 and 2025 in F555W and F467M filters (see Fig.~\ref{fig:fov}). Because of the brightness of Proxima, these images are very short so as not to saturate the target. Unavoidably, the setup of the F555W observations (which were also taken in subarray mode, with only one chip used) limits the number (less than 10 in the worst case) and the signal  (up to six magnitudes fainter than Proxima) of the surrounding sources. Given the relatively redder color of Proxima with respect to the other field stars, the choice of the F467M filter in the latest programs leads to a less-extreme exposure time for the observations, which results in a less dramatic scene (20--30 reference stars, only three magnitudes fainter than Proxima). We will discuss in the next Section how these different setups impact the astrometric registration of our images.\looseness=-4

We analyzed calibrated (dark-/bias-subtracted, flat-fielded) \texttt{\_flc} exposures corrected for charge-transfer-efficiency (CTE) defects, which affects positions and flux of sources in the detector, as described in \citet{2010PASP..122.1035A,2018acs..rept....4A}. The \texttt{\_flc} images are also best suited for astrometry because they are not resampled, thus retaining the real flux distribution of the sources that would be washed out by the mosaicking process in drizzled images \citep{2022acs..rept....2A}.\looseness=-4

The region of sky currently containing Proxima is sparse, so there is no need for a two-step astro-photometric reduction as in crowded fields \citep[e.g.,][]{2018LibralatoNGC362,2022LibralatoPMcat}. We run the code \texttt{hst1pass}\footnote{\href{https://www.stsci.edu/~jayander/HST1PASS/}{https://www.stsci.edu/~jayander/HST1PASS/}} \citep{2022acs..rept....2A} to measure position and flux of all detectable sources in the field by means of point-spread-function (PSF) fitting. We made use of effective-PSF (ePSF) models obtained by fine-tuning (for each image) publicly-available sets of library ePSFs derived as in \citet{2017MNRAS.470..948A}. Positions were corrected for geometric-distortion (GD) effects using the WFC3/UVIS GD corrections by \citet{2009BelliniWFC3,2011BelliniWFC3}. The GD of a detector changes over time, even for the instruments onboard of \hst, although at a slower pace. In the next section, we describe an additional fine-tuning of the GD correction to compensate for these changes, which primarily impacted the more-recent F467M images. Finally, the astro-photometric catalog of each exposure contains a parameter called \qfit, which represents the goodness of the ePSF fit \citep{2014LibralatoHAWKI,2022acs..rept....2A}. Bright, well-measured sources have a \qfit close to 0, whereas poorly-measured and/or faint objects have a larger \qfit value. We use this parameter to discard low-quality Proxima measurements later in the astrometric-fit stage.\looseness=-4

\section{Determination of the astrometric parameters}\label{sec:astrometry}

\citet{2017BedinLuhman,2024BedinLuhman} and \citet{2018MNRAS.481.5339B,2020MNRAS.494.2068B} have detailed a procedure that makes use of the \gaia catalog to set up a common reference frame to compare multi-epoch observations of a target and fit its astrometric parameters. We followed a similar approach, but implemented some changes and improvements.

\subsection{Astrometric reference frame and GD refinement}\label{sec:refframe}

We used the \gaia DR3 catalog to set up a common reference frame for the \hst data of Proxima. We considered only the \gaia sources in our field that fulfill all the following conditions:\looseness=-4

~\\ 
\schema[open]{}{
\noindent\texttt{source\_id} $\ne$ 5853498713190526720 ,\\
\noindent$13 \le G \le 20$ ,\\
Renormalized unit weight error \texttt{RUWE} $<$ 1.4 ,\\
\texttt{parallax} $>$ 0 mas ,\\
\texttt{parallax\_error} $<$ 0.25 mas ,\\
0 $<$ \texttt{pmra\_error} $<$ 0.25 \masyr ,\\
0 $<$ \texttt{pmdec\_error} $<$ 0.25 \masyr ,\\
\texttt{astrometric\_params\_solved} $>$ 3.
}
\\
~\\
\noindent The first condition excludes Proxima itself from the sample. Parallaxes were corrected for the zero-point bias described in \citet{2021gaiaplxbias} using the associated code\footnote{\href{https://gitlab.com/icc-ub/public/gaiadr3_zeropoint}{https://gitlab.com/icc-ub/public/gaiadr3\_zeropoint}}. This correction works only for sources with a five- or six-parameter solution, a requirement which is ensured by the last condition.\looseness=-4

For each image, we propagated the \gaia positions in Equatorial coordinates to the epoch of the observation. This epoch propagation was performed using the tool \texttt{PyGaia}\footnote{\href{https://github.com/agabrown/PyGaia/tree/master}{https://github.com/agabrown/PyGaia/tree/master}}, which is designed for \gaia science performance simulation and data manipulation of astrometric catalogs. The repository documentation warns users that this package is not intended for accurate astrometry applications, and that the \texttt{Astropy} \citep{astropy:2013,astropy:2018} facility should be used instead. The \texttt{PyGaia} epoch-propagation routines follow the prescriptions in the official \gaia DR3 documentation \citep{2022gdr3.reptE...4H}, whereas those in \texttt{Astropy} include pieces of software from the International Astronomical Union's Standard of Fundamental Astronomy (SOFA) repository\footnote{\href{http://www.iausofa.org}{http://www.iausofa.org}}. However, the former tool performs much faster than the latter. We compared a set of simulated, epoch-propagated positions of a Proxima-like target (high parallax and PM; known LOS velocity) and found that within the temporal baseline covered by our \hst observations, the coordinates computed with the two tools differ by 10 $\mu$as at most in 20 years, well below the accuracy of our \hst data. For this reason, we used \texttt{PyGaia} in our work.\looseness=-4

The aforementioned epoch propagation was performed with a linear model (``standard model of stellar motion''), starting from the assumption that stars are moving with uniform velocity relative to the Solar-System Barycenter. While this model includes effects like the perspective acceleration, which can be significant for high-PM stars close to the Sun like Proxima, it cannot be directly compared with our observations that are taken from (close to) the Earth at different times of the year, thus including the effect of the annual parallax. We modeled the annual-parallax contribution by adding to the position of the barycenter of the Solar System at a given epoch the following quantity:
\begin{equation*}\label{eq1}
    \left[\begin{array}{cc}
    \Delta \alpha & \Delta \delta
    \end{array}\right] = 
    \left[\begin{array}{c}
    {\rm p_\alpha} \\ {\rm p_\delta}
    \end{array}\right] \cdot \varpi \, ,
\end{equation*}
where $\varpi$ is the parallax and $({\rm p}_\alpha, {\rm p}_\delta)$ denote the parallactic factors computed using the \texttt{python} version of the Naval Observatory Vector Astrometry Software \citep[NOVAS][]{2011AAS...21734414B} package\footnote{\href{https://pypi.org/project/novas/}{https://pypi.org/project/novas/}}.\looseness=-4

These epoch-propagated Equatorial coordinates were then projected on to a tangent plane centered at $(\alpha,\delta)_{\rm tan} = (217.392321472, -62.676075117)$, i.e., the position of Proxima in the \gaia DR3 catalog (epoch 2016.0), using, e.g., Eqs.~(3) and (4) in \citet{2018MNRAS.481.5339B}. These projected coordinates remain in angular units (e.g., degrees), which makes direct comparison with the pixel-based coordinates in the \hst-based catalogs difficult. Therefore, we applied three changes: (i) we divided these projected positions by an arbitrary pixel scale (40 \maspixel, which is similar to the pixel scale of the WFC3/UVIS camera); (ii) we flipped the $x$ direction to align the detector's $x$ axis with the common convention of pointing towards West (the $y$ axis is oriented towards North); (iii) we added an offset of $\Delta (x,y) = (5000,5000)$ pixels to avoid working with negative coordinates. All these choices are completely arbitrary, but are necessary to setup a common, pixel-based reference frame that will allow us to transform positions in Equatorial coordinates to the pixel-based coordinate system of any given \hst image, and vice-versa, in the next part of the work. We stress that the actual values used for shift/scale/orientation do not matter, what is important is their consistency throughout the entire process.\looseness=-4

We cross-identified the \hst-based catalog of each image with the corresponding \gaia-based catalog prepared as just described. We collected all \hst-\gaia position pairs obtained this way, and used them to refine the GD correction of the WFC3/UVIS detector following the prescriptions of \citet{2009BelliniWFC3}. Any GD correction is derived by comparing a set of observed positions (taken with the detector used to solve for the GD) and a set of distortion-free positions. In our case, the former set corresponds to the \hst-based positions, whereas the latter set is provided by the \gaia catalog \citep[see, e.g.,][for other works that used the \gaia catalog to solve for the GD correction of an imager]{2023GriggioNIRCam,2024GriggioVVV,2024LibralatoMIRI}.

The refinement of the WFC3/UVIS GD correction can be summarized as follows:
\begin{itemize}
    \item for each chip, we used four-parameters linear transformations (two offsets, change of global scale, rotation) to transform the \gaia-based positions on to the GD-corrected reference frame of the corresponding \hst catalog. The coefficients of such transformations were computed by means of the positions of bright, well-measured stars in both catalogs;
    \item we applied an inverse GD correction to get these transformed \gaia positions in the raw \hst reference system;
    \item we defined the positional residuals as the difference between the \hst-based and the \gaia-based raw positions;
    \item we modeled these positional residuals \citep[normalized as in][]{2009BelliniWFC3} with two third-order polynomial functions, which represent the additional GD correction for the WFC3/UVIS detector;
    \item we added the new GD correction to the original one from \citet{2009BelliniWFC3,2011BelliniWFC3}.
\end{itemize}
We iterated the entire process (cross-match and GD correction) 10 times, after which we found negligible differences between the polynomial coefficients in two consecutive iterations. The GD refinement for the F555W solution does not provide a significant change, as expected since most of the F555W data was taken a few years after the WFC3/UVIS installation so that the GD correction from \citet{2009BelliniWFC3,2011BelliniWFC3} is still optimal. For the F467M filter, we found that the GD has changed over time, with small trends as a function of the $y$ position. After the correction, this GD systematic has disappeared.\looseness=-4

\subsection{The catalog of Proxima's positions}\label{sec:catalog}

The next step consists in creating the catalog with the multi-epoch measurements of the position of Proxima. While Proxima's coordinates in every exposure come directly from the ePSF fit described in Sect.~\ref{sec:datared}, there are other quantities required prior to move to the next stage: a pixel-to-Equatorial coordinate transformation (and its inverse), and an estimate of the positional error of Proxima.\looseness=-4

For each, image we cross-identified the same stars and computed the six-parameters linear transformations (similar to the four-parameter transformations with the addition of two so-called skew terms that described the change of scale between $x$ and $y$ axes and the change of perpendicularity between $x$ and $y$ axes) necessary to transform positions as measured in the \hst reference frame on to that of \gaia (defined as in Sect.~\ref{sec:refframe}). The coefficients of such transformations were computed by means of the positions of bright, well-measured stars in both catalogs (Proxima excluded). We also ensured that these stars lay in the same WFC3/UVIS chip as Proxima. As shown in many astrometric works with \hst data \citep[e.g.][]{2014ApJ...797..115B,2018LibralatoNGC362,2022LibralatoPMcat}, a local approach works best in minimizing residual systematics that can affect astrometry. Using stars within the same chip is an appropriate compromise between locality and statistics. Together with the details provided in Sect.~\ref{sec:refframe}, these transformations contain the necessary information to transform positions between the \hst- and \gaia-based pixel reference frame, as well as to the ICRS (in Equatorial coordinates), at any given epoch/image.\looseness=-4

For the positional-error budget of each estimate of Proxima's position, we need to include two contributions: (i) the positional error of Proxima; and (ii) the error of the aforementioned transformations. To estimate the former value, we used the empirical relation between positional error and magnitude described in Sect.~5.2 and Fig.~2 of \citet[provided via private communication by the first author]{2014ApJ...797..115B}. Proxima is very bright (instrumental magnitude $\lesssim$$-13.5$) in all our exposures and the predicted positional error is $<$0.01 WFC3/UVIS pixel ($<$0.4 mas), in agreement with the expected positional precision for bright stars in \citet{2011BelliniWFC3}.\looseness=-4

The latter contribution is dominated by the positional error in the \hst images of the stars used to compute the coefficients of the linear transformations. We estimated this value by transforming the position of Proxima from the \hst-based to the \gaia-based pixel coordinate system using six-parameter linear transformations. The coefficients of these transformations were calculated by adding to each \hst position a random noise (drawn from a Gaussian distribution with 0 mean and $\sigma$ equal to the expected magnitude-dependent error of each star). We repeated this operation 1\,000 times and defined the error due to the transformations as the standard deviation of all these measurements. The final positional uncertainty for each Proxima measurement was defined as the sum in quadrature of these two errors. This error is, on average, 0.03 WFC3/UVIS pixel. For the F467M data, which have many stars in common between \hst and \gaia measured with a relatively-high signal-to-noise ratio, all measurements have about the same positional uncertainty (0.03 WFC3/UVIS pixel). For the F555W data, which can have less than 10 stars in common and with a three-to-six magnitudes of difference with Proxima, the uncertainty ranges between 0.02 and 0.06 WFC3/UVIS pixel.\looseness=-4

Finally, before determining the astrometric parameters, we analyzed the catalog of Proxima to identify entries with poor quality, looking for each filter at the positional-uncertainty and \qfit values as a function of magnitude. While the F467M data looked consistent, the F555W data showed clear outliers in the distribution. Thus, we excluded all measurements with a positional uncertainty larger than 0.05 WFC3/UVIS pixel (2 mas), and \qfit value larger than 0.07. The latter threshold was defined by visually inspecting the distribution of the \qfit as a function of instrumental magnitude. We noticed a skewed distribution of the \qfit values with average 0.05 and $\sigma \sim 0.02$. We chose a conservative cut at 0.07. The \qfit selection is applied only in the direction of increasing \qfit and not the opposite direction (i.e., mean $-1\sigma$) because a \qfit $<0.05$ is usually considered the proxy of a well-measured star. In total, the final sample for the astrometric analysis comprised 59 two-dimensional measurements.\looseness=-4

\begin{figure*}[t]
    \centering
    \includegraphics[width=\textwidth]{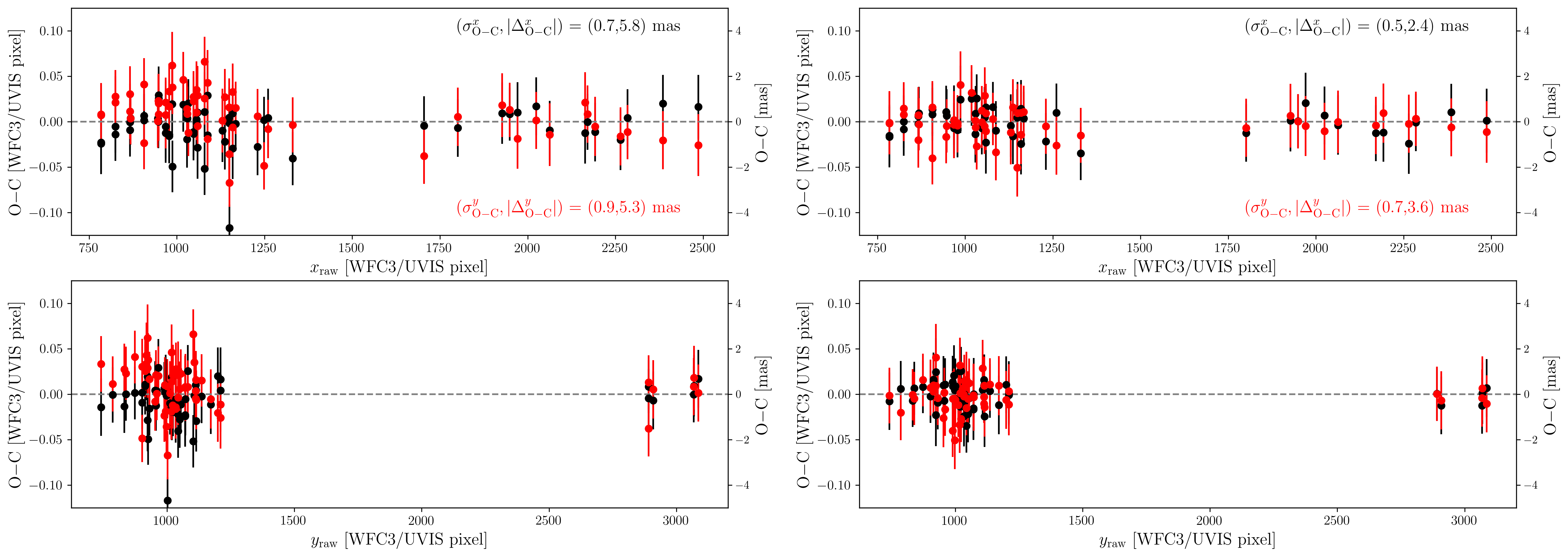}
    \caption{Positional residuals from the model (O$-$C) as a function of $x_{\rm raw}$ and $y_{\rm raw}$, respectively from top to bottom. Black points refer to the $x$ residuals, while red points represent the $y$ residuals. The gray dashed line in each plot is set at 0 for reference. The left panels show the O$-$C at the first iteration of the astrometric fit, while the right panels display the residuals at the last iteration of the fit (less points are visible because of the iterative rejection of outliers, see the text for details). In the top panels, we report the standard deviation ($\sigma_{\rm O-C}$ and the absolute maximum scatter of the points ($|\Delta_{\rm O-C}|$) as a reference.\looseness=-4}
    \label{fig:xyres}
\end{figure*}

\subsection{Determination of the astrometric parameters}\label{sec:astromodel}

The goal of this paper is to determine the astrometric parameters of Proxima at a specific epoch, namely: the Equatorial coordinates $(\alpha_{\rm ref}, \delta_{\rm ref})$, the PM $(\mu_\alpha \cos\delta, \mu_\delta)$, and the parallax $\varpi$. We chose $t_{\rm ref} = 2016.0$, so to directly compare our findings with those in the \gaia DR3 catalog.\looseness=-4

We derived these astrometric parameters by comparing our \hst measurements with a model representing the expected path of Proxima in the plane of the sky. The model was obtained by combining a linear motion (from \texttt{PyGaia}) and an annual-parallax displacement (using the NOVAS package as described in Sect.~\ref{sec:refframe}). A more sophisticated model for the stellar motion, as that provided by \texttt{PyGaia}, is needed to include the effect of the perspective acceleration. In our work, we did not use any LOS RV velocity measurement in the fit, but assumed a constant value of $-21.94$ km s$^{-1}$ from the \gaia DR3 catalog.\looseness=-4

Proxima differs significantly in both color and magnitude from the adopted network of reference stars used for the alignment, i.e., sources in the \gaia DR3 catalog with well-determined PMs and parallaxes. Although using \texttt{\_flc} exposures should remove most systematics associated to the magnitude dependence of the CTE (and we will describe later how residual CTE systematic errors are also removed in the process), additional residuals may still arise owing the color difference of the target. In wide filter passbands, the exact spectral energy distribution (SED) of each star affects both the fine structure of the PSF and the GD of the camera via optical refraction \citep[see, e.g.,][for an example of the relation between GD correction and stars' SED]{2011BelliniWFC3}. These color-dependent effects are equivalent to a form of differential chromatic refraction. We investigated the possible contribution of these effects, finding to be small for Proxima. Nevertheless, we report the results of the astrometric fit with a possible chromatic correction in Appendix~\ref{appendix:fitcolcor}.\looseness=-4

We developed our suits for the astrometric fit so that the comparison between observations and models happens in the image plane (specifically, the GD-corrected reference frame of each \hst image) instead of the sky plane, in preparation for the analysis of the trailed images in future works. The implementation of this step is trivial because of how we set up our framework in Sect.~\ref{sec:astrometry}, and it just requires to: (i) project the sky coordinates predicted by the model at any given epoch using the tangent point $(\alpha,\delta)_{\rm tan}$ provided in Sect.~\ref{sec:refframe}; and (ii) transform each model position from the projected sky plane to the corresponding image plane of the \hst exposure using the six-parameter linear transformations described in Sect.~\ref{sec:catalog}.\looseness=-4

\begin{table*}[t]
\centering
\caption{Fitted astrometric parameters for Proxima.}
\begin{tabular}{llccc}
\hline
Parameter & Prior & Best-fit value & \gaia DR3 value & Agreement in $\sigma$\\
\hline
$\alpha_{\rm 2016.0}$ [deg]             & $\mathcal{U}(217.35, 217.41)$ & $217.392321386$ $\left(_{-0.386}^{+0.422}\,{\rm mas}\right)$ & $217.392321472$ $\left(_{-0.024}^{+0.024}\,{\rm mas}\right)$ & 0.54 \\
$\delta_{\rm 2016.0}$ [deg]             & $\mathcal{U}(-62.69,-62.65)$  & $-62.676075116$ $\left(_{-0.187}^{+0.186}\,{\rm mas}\right)$ & $-62.676075117$ $\left(_{-0.034}^{+0.034}\,{\rm mas}\right)$ & 0.01 \\
$\mu_\alpha \cos\delta$ [mas yr$^{-1}$] & $\mathcal{U}(-4000,-3500)$    & $-3781.761$     $\left(_{-0.050}^{+0.045}\right)$            & $-3781.741$     $\left(_{-0.031}^{+0.031}\right)$            & 0.28 \\
$\mu_\delta$  [mas yr$^{-1}$]           & $\mathcal{U}(700,800)$        & $769.458$       $\left(_{-0.047}^{+0.048}\right)$            & $769.465$       $\left(_{-0.051}^{+0.051}\right)$            & 0.08 \\
$\varpi$ [mas]                          & $\mathcal{U}(700,800)$        & $768.373$       $\left(_{-0.162}^{+0.194}\right)$            & $768.067$       $\left(_{-0.050}^{+0.050}\right)$            & 1.19 \\
\hline
\end{tabular}
\tablefoot{The second column provides the prior used in the MCMC fit. The third column contains the best-fit parameters (and $\pm1\sigma$ uncertainty). The last two columns report the corresponding value in the \gaia DR3 catalog, and the agreements in $\sigma$ with our \hst-based measurement.\looseness=-4}
\label{tab:astrofit}
\end{table*}

The best-fit astrometric parameters and corresponding errors were obtained by using the affine-invariant Markov Chain Monte Carlo (MCMC) method \texttt{emcee} \citep{2013PASP..125..306F}. For each parameter, we defined an initial guess using a value close to the Proxima entry in the \gaia DR3 catalog, and then allowed it to vary uniformly within a wide range of values (see Table~\ref{tab:astrofit}). Starting from the initial guesses, we initialized 100 walkers and let them evolve for 5000 steps, maximizing the log-likelihood defined as:
\begin{align*}
    \ln \mathcal{L} &= - \frac{1}{2} \left[\chi_{x}^2 + \chi_{y}^2 + \sum_{i = 1}^N{\ln(2\pi\sigma_x^2)} + \sum_{i = 1}^N{\ln(2\pi\sigma_y^2)} \right] \, ,
\end{align*}
where $N$ is the number of measurements, $(\sigma_x,\sigma_y)$ are the positional errors defined in Sect.~\ref{sec:astrometry}, and:
\begin{align*}
    \chi^2 &= \sum_{i = 1}^N \left(\frac{O_i - M_i}{{\sigma_O}_i}\right)^2 \, ,
\end{align*}
with $O_i$, ${\sigma_O}_i$ and $M_i$ are the observed data-point, the observed uncertainty and the model values (either $x$ or $y$ position), respectively. The best-fit solution was defined the parameter set that maximizes the log-probability, after removing the first 20\% of the steps as burn-in phase. The corresponding errors were computed as the 16-th and 84-th percentiles about the best-fit solution of the posterior distributions.\looseness=-4

Working on the $x$/$y$ image plane allows us to search for systematic trends in the data. The observation-model (O$-$C) residuals should be, on average, zero regardless the position on the detector where Proxima was imaged. After the first iteration, we noticed that this requirement was not fulfilled, meaning that some unknown systematics and/or the contribution of outliers are affecting our fit. The left-most panels of Figure~\ref{fig:xyres} shows an overview of the (O$-$C) $x$ and $y$ residuals at the first iteration of the astrometric fit as a function of $(x_{\rm raw},y_{\rm raw})$ position on the detector. The first iteration produced a solution with a scatter that can be as large as 5.8 mas, ($\sim$0.14 WFC3/UVIS pixel), which seem to correlate with the position on the detector. These systematics could be related to residual CTE defects (which can be important for the F555W images since their exposure time is 1\,s), as well as to chromatic/magnitude effects in the ePSF (Proxima is redder and 3--6 mag brighter than any other source in the field). We corrected these systematics by fitting a linear relation to the (O$-$C) residuals as a function of $x_{\rm raw}/y_{\rm raw}$ as done by \citet{2017BedinLuhman}. This correction was computed separately for each filter.\looseness=-4

\begin{figure*}[t]
    \sidecaption
    \centering
    \includegraphics[width=12.9cm]{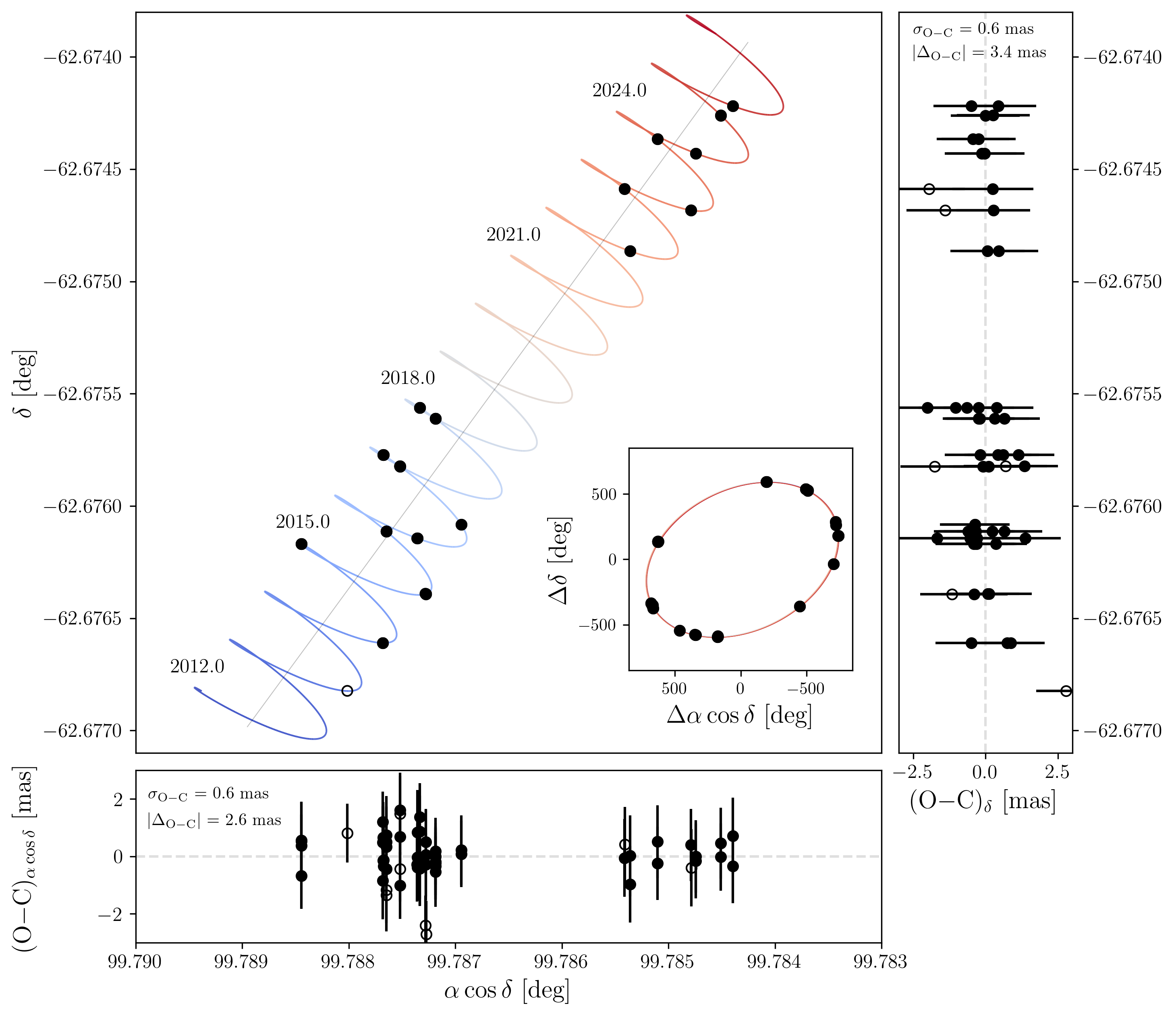}
    \caption{Series of \hst observations of Proxima. The main panel shows the path in the $(\alpha\cos\delta,\delta)$ plane of Proxima. The filled black dots are the \hst measurements used in the last iteration of the astrometric fit, while the open dots refer to the rejected measurements. The gray, solid line is the expected linear motion (i.e., without the annual-parallax contribution) of Proxima, whereas the colored line represents the astrometric solution of Proxima. The line color changes from blue to white to red as the epoch increases (from the bottom left to the top left of the plot). The inset in the main panel shows the residuals from the linear motion, which highlight the annual-parallax ellipse. The residuals of the astrometric fit as a function of $\alpha\cos\delta$/$\delta$ coordinates are displayed in the side panels. The gray, dashed line is set to 0 as a reference. In each of these side panels, we report the standard deviation ($\sigma_{\rm O-C}$ and the absolute maximum scatter of the points ($|\Delta_{\rm O-C}|$).\looseness=-4}
    \label{fig:sky}
\end{figure*}

However, in contrast with \citet{2017BedinLuhman} where the dataset used was homogeneous (multi-epoch data taken with the same instrument/filter/exposure time) and observations were designed ad-hoc for their target, for Proxima we have two distinct datasets. In particular, the 1-s exposures in the F555W filter are the most problematic: there are a dozen stars in the field of Proxima in the best case, so the image alignment (Sect.~\ref{sec:astrometry}) was not as robust as for the F467M filter, resulting in a larger scatter in the (O$-$C) residuals for the F555W data than for the F467M one. Such scatter made it difficult to remove bad measurements before the astrometric fit. For this reason, after each iteration we flagged and excluded all measurements that for which the (O$-$C) $x$/$y$ residuals from the previous iteration were more than 3$\sigma$ away from the median value of the residual distribution.\looseness=-4

We iterated the entire procedure (astrometric fit, systematic correction, outlier rejection) three times, after which we found no clear systematics in the (O$-$C) residuals (see rightmost panels of Fig.~\ref{fig:xyres}) and the number of outliers did not change in two consecutive iterations (12 out of 59 data points). We estimated the goodness of our fit by computing the reduced chi-square $\chi^2_{\rm r} = \chi^2 / {\rm dof}$, where ``dof'' means degree of freedom, i.e., the difference between the number of data points and the number of fitting parameters. The final $\chi^2_{\rm r}$ is 0.41. The best-fit astrometric solution
\footnote{Proxima passed close to faint background sources around epochs 2014.8 and 2016.2, causing microlensing events that shifted these star positions \citep{2014ApJ...782...89S, 2018MNRAS.480..236Z}. These sources are not included in the network of objects used to setup our reference frame (Sect.~\ref{sec:refframe}) because they are not visible in the F555W data ($>$8.5 mag fainter than Proxima), so we anticipate no impact on our analysis. To confirm this, we re-measured Proxima's astrometric parameters excluding the images taken near these events. The parameters were consistent with those from the full sample at the 0.5$\sigma$ level, except for the parallax, which showed consistency at $\sim$2$\sigma$. The parallax changed the most because the excluded 2014.8 images are near the maximum parallax elongation, making them crucial for constraining that parameter.\looseness=-4}
is reported in Table~\ref{tab:astrofit}, and the corner plots are provide in Appendix~\ref{appendix:corner}. A visual representation of the astrometric solution of Proxima, together with the (O$-$C) residuals along $(\alpha\cos\delta,\delta)$ directions, is presented in Fig.~\ref{fig:sky}.
Table~\ref{tab:astrofit} also reports the values taken from the \gaia DR3 catalog for comparison. The agreement between the two sets of parameters is at the 1-$\sigma$ level in the worst case, with the largest discrepancies being $\Delta\alpha_{2016} \sim 0.3$ mas and $\Delta\varpi \sim 0.3$ mas. Remarkably, our and \gaia PMs are in agreement within 20 $\mu$as yr$^{-1}$. Our positional and parallax uncertainties are larger than \gaia's, especially for $\alpha_{2016.0}$, but in agreement with the typical positional uncertainty of a bright star with \hst \citep[0.4 mas $\sim$ 0.01 WFC3/UVIS pixel;][]{2009BelliniWFC3,2011BelliniWFC3}. Appendix~\ref{appendix:positions} provides the positions of Proxima used in our analysis.\looseness=-4

\section{Proper motion anomaly using \hipparcos, \gaia and \hst}\label{sec:pma}

Our astrometric time series can potentially be used to search for residual astrometric signatures in the (O$-$C) plane. However, the precision achieved by our point-source images (that can be as large as 1 mas) is comparable to the amplitude of the residuals we want to investigate, making it challenging to draw any conclusion about potential signatures with enough statistical confidence. An alternative method to search for an unseen companion, commonly used when dedicated astrometric series are not available, is the so-called PM anomaly (PMa).\looseness=-4

If a star is part of a binary system, its apparent motion in the plane of the sky varies over time due to the gravitational influence of its companion(s). The amplitude of this perturbation depends on the mass of the companion and the overall orbital configuration (e.g., the semi-major axis). The PMa technique aims at detecting these deviations by comparing two independent sets of PMs for the source: the long-term PM (in which orbital perturbations average out) and short-term PM (an ``instantaneous'' measurement of the star's motion). \citet{2019KervellaPMa} took advantage of the \hipparcos and \gaia DR2 catalogs to measure this PMa by comparing the \gaia DR2 (short-term) PM and the \hipparcos-\gaia (\textit{Hip}-\gaia; long-term) PM. Later, \citet{2022KervellaPMa} updated the PMa values with the \gaia DR3 catalog. We followed the prescriptions \citet{2019KervellaPMa,2022KervellaPMa} to compute the PMa by combining our \hst long-term PM (temporal baseline of 10.89 yr) with the short-term PM provided by the \gaia DR3, and try to infer the mass of Proxima\,c, the only putative exoplanet around Proxima for which the PMa method is sensitive.\looseness=-4

First, we measured the PMa vector by combining our \hst and the \gaia DR3 PMs. This computation has to be done in three dimensions (analogous to the ``standard model of linear motion'' introduced in Sect.~\ref{sec:astromodel}) to take into account effects like the parallax and perspective acceleration of Proxima. Our \hst-\gaia PMa for Proxima is:
\begin{equation*}
    \Delta\mu_{HST-Gaia} = (0.056 \pm 0.056, -0.009 \pm -0.068) \textrm{ mas yr}^{-1} \, ,
\end{equation*}
which is similar to what found by \citet{2022KervellaPMa}:
\begin{equation*}
    \Delta\mu_{Hip-Gaia} = (-0.022 \pm 0.046, -0.069 \pm 0.069) \textrm{ mas yr}^{-1} \, .
\end{equation*}
Both estimates present a marginal significance.\looseness=-4

To derive a mass for Proxima\,c from the PMa, we followed the procedure described in \citet{2019KervellaPMa}. This methodology assumes a simplified geometry (a binary system with circular orbit seen with a face-on orientation), meaning that more complex orbital configurations can lead to different solutions. Upon these assumptions, the mass of the companion ($m_2$) scales with the square root of the mass of the primary star ($m_1$) and the observed PMa ($\Delta\mu$), but it is also degenerate with orbital radius $r$:
\begin{equation*}
    \frac{m_2}{\sqrt{r}} = \sqrt{\frac{m_1}{G}} \cdot \left( \frac{\Delta\mu \textrm{ [mas yr}^{-1}\textrm{]}}{\varpi \textrm{ [mas]}}\cdot 4740.470\right) \, ,
\end{equation*}
where $G$ is the universal gravitational constant, and 4740.470 is a conversion factor from \masyr to m s$^{-1}$. \citet{2019KervellaPMa} showed that additional factors should be included: the orbital inclination, the time-window smearing of the short-term PM, and the sensitivity of the long-term PM. The first term was defined in a statistical fashion as in Sect. 3.6.1 of \citet{2019KervellaPMa}. The smearing factor, necessary to compensate for the fact the short-term PM is not instantaneous but taken over an extended period of time (1038 days for \gaia DR3), was computed as in Sect. 3.6.2 of \citet{2019KervellaPMa}. The factor related to the sensitivity of the long-term PM is included so not to bias the PMa vector if the orbital period of the binary system is significantly longer than the long-term-PM temporal baseline. In our work, we assumed the results of the simulations in Sect. 3.6.3 of \citet{2019KervellaPMa} and shown in their Fig.~3, rescaled by our long-term temporal baseline.\looseness=-4

\begin{figure*}[t!]
    \sidecaption
    \centering
    \includegraphics[width=12.9cm]{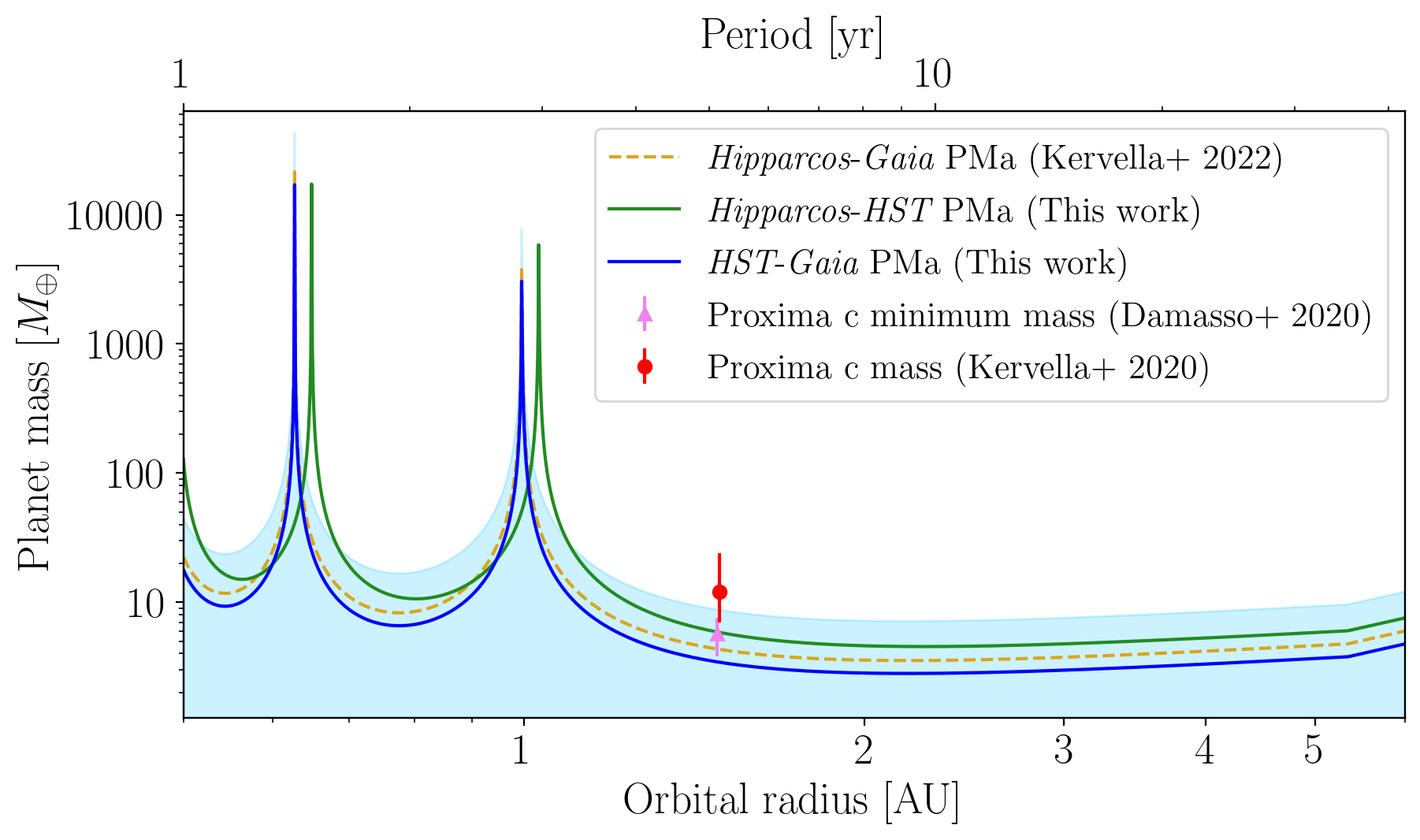}
    \caption{Expected mass of Proxima~c as a function of its orbital radius derived from the PMa analysis. The solid blue line shows the mass estimate obtained in this work from the combination of \hst and \gaia-DR3 astrometry, with the shaded light-blue region representing the $1\sigma$ uncertainty. The relation obtained combining \textit{Hip}-\gaia (just ``\hipparcos'' in the plot legend for clarity) and part of the \hst data is shown as a solid green line. The dashed yellow line represents the PMa-based relation obtained by \citet{2022KervellaPMa} using \hipparcos and \gaia. The pink triangle marks the minimum mass derived of Proxima~c from \citet{2020SciA....6.7467D}, while the red point shows the mass estimate from \citet{2020KervellaProxima}.}
    \label{fig:pma}
\end{figure*}

Figure~\ref{fig:pma} shows the expected mass of Proxima\,c as a function of the orbital radius, assuming a mass of 0.1221 $M_\odot$ for Proxima \citep{2015ApJ...804...64M}. The blue line represents our solution, while the light-blue region sets the $1\sigma$ uncertainty on the mass at any given radius, which was computed including the contributions of the PM errors with a Monte Carlo approach. The dashed gold line is the relation derived using the PMa from \citet[the uncertainty region is omitted for clarity]{2022KervellaPMa}. The spikes at radius lower than 1 AU mark the parameter space in which our PMa is not sensitive. The pink triangle is the minimum mass of Proxima\,c from the work of \citet[$m_{\rm c} \sin i = 5.7 \pm 1.9$ $M_\oplus$]{2020SciA....6.7467D}, while the red point represents the mass measured by \citet{2020KervellaProxima}, which used the PMa of \citet{2019KervellaPMa} and the results from \citet{2020SciA....6.7467D} to estimate the orbital inclination  ($i = 152 \pm 14$ deg) and mass of Proxima\,c ($m_{\rm c} = 12^{+12}_{-5}$ $M_\oplus$).

Assuming a circular orbit for Proxima\,c with radius equal to the semi-major axis $a = 1.48$ AU from \citet{2020SciA....6.7467D}, we obtain a mass for Proxima\,c of $m_c = 3.4^{+5.2}_{-3.4}$ $M_\oplus$.  A similar result is obtained considering the PMa from \citet[$m_c = 4.2^{+4.9}_{-4.2}$ $M_\oplus$]{2022KervellaPMa}.\looseness=-4

We also run the analysis using the \textit{Hip}-\gaia long-term PM from \citet{2022KervellaPMa} and a short-term PM obtained using part of the \hst data at our disposal (green line in Fig.~\ref{fig:pma}). Specifically, we re-ran our analysis using only the F467M data taken in 2020s. We fixed $\alpha_{2016}$, $\delta_{2016}$ and $\varpi$ to the values obtained in Sect.~\ref{sec:astromodel} and just fitted the PM components. These data span a temporal baseline of $\sim$3 yr, similar to that of the \gaia DR3. We find a PM of:
\begin{align*}
    \mu_\alpha \cos\delta & = -3781.754_{-0.046}^{+0.048} \text{ mas yr}^{-1} \, ,\\
    \mu_\delta            & =   769.459_{-0.047}^{+0.046} \text{ mas yr}^{-1} \, ,  
\end{align*}
which leads to:
\begin{align*}
    \Delta\mu_{Hip-HST} & = (-0.071 \pm 0.057, -0.059 \pm 0.065) \textrm{ mas yr}^{-1} \, .
\end{align*}
Using this value for the PMa, we obtain $m_{\rm c} = 5.9^{+5.5}_{-5.5}$ $M_\oplus$. Although all estimates are in agreement at 1$\sigma$ level with the minimum mass from \citet{2020SciA....6.7467D}, these PMa-based masses of Proxima\,c are likely biased by a mix of large PM uncertainty and smearing effect of the \gaia DR3 or \hst short-term PMs. It is worth noting that recent studies using LOS RVs \citep[e.g.,][]{2022A&A...658A.115F,2025A&A...700A..11S} have not found clear signatures of Proxima\,c, further complicating the puzzle. Additional astrometric monitoring is needed to constrain the existence and mass of Proxima\,c.\looseness=-4

\section{Conclusions}\label{sec:conclusions}

This first paper of the ARES project presents a refined astrometric analysis of Proxima using multi-epoch \hst WFC3/UVIS observations obtained over 13 years. By improving the GD correction of the WFC3/UVIS detector and combining \hst and \gaia-DR3 astrometry, we achieved sub-milliarcsecond precision on Proxima's position, PM, and parallax. Our best-fit solution agrees with that from \gaia DR3 within $\sim$1$\sigma$, validating the methodology and error budget. Our measurement also provides an independent test for the \gaia's astrometry of this very-red M dwarf. Indeed, we linked our astrometry to that of \gaia using all suitable stars in the field except Proxima, and then fit Proxima’s position, PM and parallax. If our \hst or \gaia measurements of Proxima were affected by a color-dependent systematic that made Proxima ``behave'' differently from the other bluer sources, we should have found a discrepancy in the astrometric parameters of Proxima determined using \hst and \gaia. Since we found no such discrepancy, we conclude that Proxima measurement is robust.\looseness=-4

We investigated the presence of the proposed exoplanet Proxima~c by measuring the PMa from the combination of \hst and \gaia PMs. The PMa method starts from the assumption of a circular and face-on orbit, so the derived mass can be different from the real one depending on the actual planetary architecture. Nevertheless, this method makes it possible to infer the mass of a companion when astrometric time series are not available. We derived a mass estimate of $m_c = 3.4^{+5.2}_{-3.4}$ $M_\oplus$ for an orbital radius of 1.48 AU (i.e., the semi-major axis of Proxima~c from past LOS RV analyses), consistent within uncertainties to RV-based measurements, but still limited by the current precision of the PMa method and data set at our disposal.\looseness=-4

This work establishes the foundations for the next phase of ARES. Future studies will exploit \hst spatial-scanning observations, capable of reaching astrometric precisions of a few tens of $\mu$as per epoch, to directly search for the signatures of low-mass companions around Proxima and other nearby systems. These upcoming results will provide a critical benchmark for validating \gaia-based discoveries and refining our understanding of planetary systems in the solar neighborhood.\looseness=-4

The next \gaia data releases will increase and improve the synergy between \hst and \gaia. \gaia's improved astrometric parameters at all magnitudes, including at the faint end, will enlarge the pool of suitable objects available to set up the reference network in studies similar to ours. Smaller uncertainties in positions, PMs and parallaxes will provide more robust epoch-propagated \gaia positions (Sect.~\ref{sec:astrometry}). However, the \hst-\gaia synergy extends beyond mutual support. \hst data can complement the \gaia astrometric series by providing additional data points and increasing the temporal baseline for the astrometric fit. Furthermore, the combination of \hst and \gaia (regardless of the data release) can yield PMs (and potentially parallaxes, depending on the temporal sampling) for sources within an \hst field that currently only have a two-parameter solution in the \gaia catalog \citep[e.g.,][]{2024A&A...692A..96L,2025arXiv251023694B}.\looseness=-4

\begin{acknowledgement}

ML acknowledges support from the ``IAF: Astrophysics Fellowships in Italy''. Based on observations with the NASA/ESA \textit{HST}, obtained at the Space Telescope Science Institute, which is operated by AURA, Inc., under NASA contract NAS 5-26555. This work has made use of data from the European Space Agency (ESA) mission {\it Gaia} (\url{https://www.cosmos.esa.int/gaia}), processed by the {\it Gaia} Data Processing and Analysis Consortium (DPAC, \url{https://www.cosmos.esa.int/web/gaia/dpac/consortium}). Funding for the DPAC has been provided by national institutions, in particular the institutions participating in the {\it Gaia} Multilateral Agreement. This research made use of \texttt{astropy}, a community-developed core \texttt{python} package for Astronomy \citep{astropy:2013, astropy:2018}, \texttt{emcee} \citep{2013PASP..125..306F}, \texttt{NOVAS} \citep{2011AAS...21734414B}, \texttt{pygtc} \citep{Bocquet2016}. This work also employed \texttt{PyGaia}, which is based on the effort by Jos de Bruijne to create and maintain the \gaia Science Performance pages (with support from David Katz, Paola Sartoretti, Francesca De Angeli, Dafydd Evans, Marco Riello, and Josep Manel Carrasco), and benefits from the suggestions and contributions by Morgan Fouesneau, Tom Callingham, John Helly, Javier Olivares, Henry Leung, Johannes Sahlmann.

\end{acknowledgement}

\bibliography{ARES_I_Proxima}{}
\bibliographystyle{aa}

\begin{appendix}

\section{Astrometric fit with chromatic-correction terms}\label{appendix:fitcolcor}

\begin{table}[t!]
\centering
\caption{Fitted astrometric parameters for Proxima with the chromatic-correction terms.\looseness=-4}
\begin{tabular}{lc}
\hline
Parameter & Best-fit value \\
\hline
$\alpha_{\rm 2016.0}$ [deg]             & $217.392321442$ $\left(_{-0.379}^{+0.393}\,{\rm mas}\right)$ \\
$\delta_{\rm 2016.0}$ [deg]             & $-62.676075111$ $\left(_{-0.150}^{+0.211}\,{\rm mas}\right)$ \\
$\mu_\alpha \cos\delta$ [mas yr$^{-1}$] & $-3781.744$     $\left(_{-0.043}^{+0.051}\right)$            \\
$\mu_\delta$  [mas yr$^{-1}$]           & $769.462$       $\left(_{-0.051}^{+0.043}\right)$            \\
$\varpi$ [mas]                          & $768.240$       $\left(_{-0.267}^{+0.180}\right)$            \\
$k_{\rm \alpha}$ [mas]                  & $0.559$         $\left(_{-0.355}^{+0.385}\right)$            \\
$k_{\rm \delta}$ [mas]                  & $0.261$         $\left(_{-0.176}^{+0.235}\right)$            \\
\hline
\end{tabular}
\label{tab:fitcolcor}
\end{table}

\begin{figure}[t!]
    \centering
    \includegraphics[width=\columnwidth]{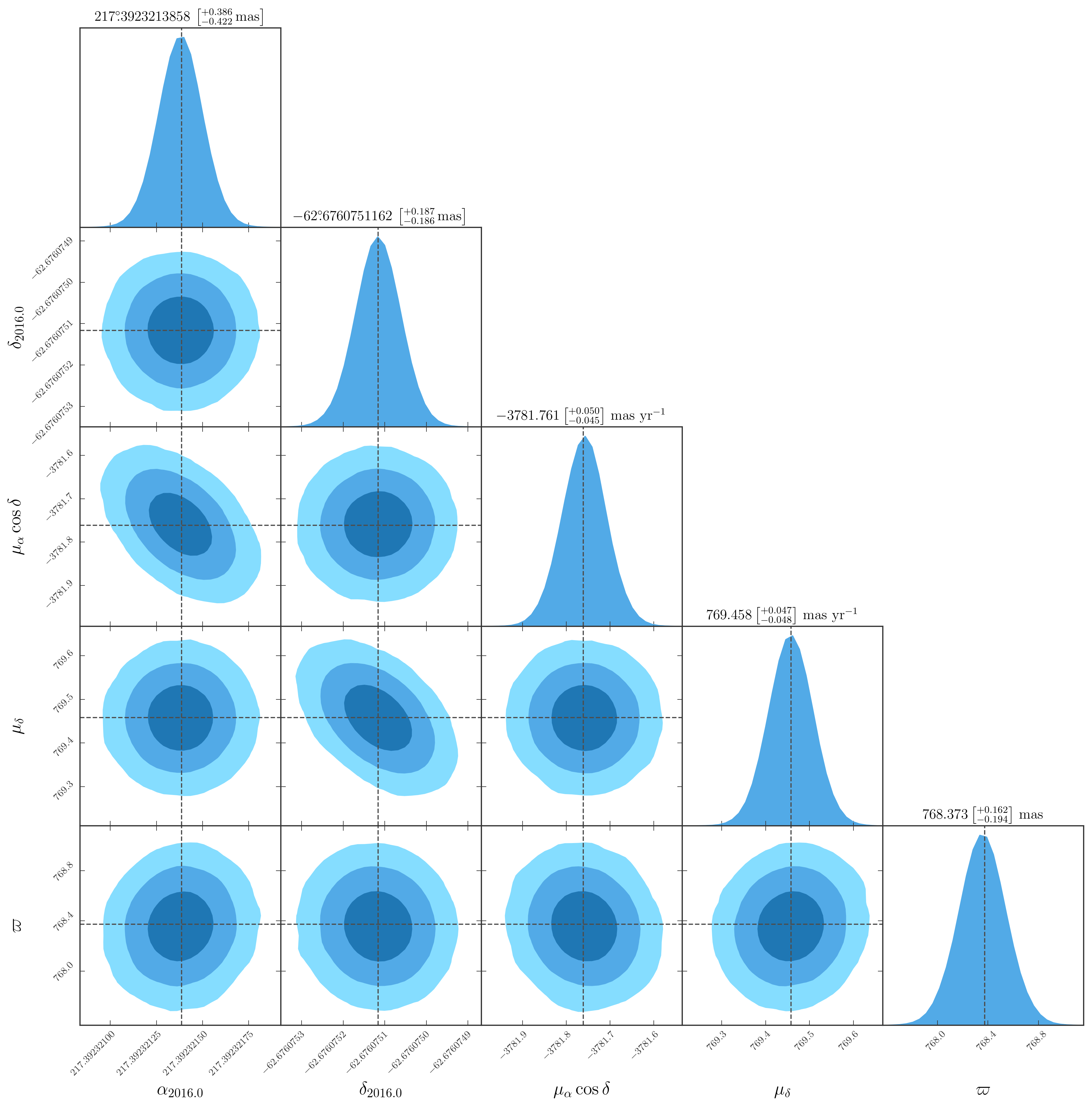}
    \caption{One- (histograms) and two-dimensional (marginalized) projections of the posterior-probability distributions of the astrometric parameters of Proxima. The dashed, gray lines set the best-fit parameters (also reported, with the corresponding uncertainties, above each histogram). The marginalized projections of the posterior-probability distributions show 1, 2 and 3 $\sigma$ contour in different shades of blue.}
    \label{fig:corner}
\end{figure}

As discussed in Sect.~\ref{sec:astromodel}, we tested the inclusion of an empirical correction. We did not attempt to disentangle the chromatic effects individually, but rather to model their combined impact on the measured centroid of the target relative to the reference star network (with a different mean color), as a function of the detector orientation of each \hst exposure with respect to the absolute reference frame set by \gaia. To this end, we introduced two additional free parameters in the astrometric fit, describing the orientation and amplitude of the corresponding systematic vector in the detector plane.
\begin{equation*}\label{eq3}
    \left[\begin{array}{cc}
    \Delta \alpha_{\rm color} & \Delta \delta_{\rm color}
    \end{array}\right] = 
    \left[\begin{array}{cc}
    k_\alpha & k_\delta
    \end{array}\right] \cdot 
    \left[\begin{array}{c}
    \frac{\sin\theta}{\cos(\delta)} \\ \cos\theta
    \end{array}\right] \, ,
\end{equation*}
where $(k_\alpha,k_\delta)$ are the amplitudes of the correction along the $(\alpha,\delta)$ directions, and $\theta$ is the position angle of the image (computed using the six-parameters linear transformations described in Sect.~\ref{sec:catalog}). The $1/\cos\delta$ term is needed to compensate for the projection effects. Accounting for this systematic contribution leads to a slightly improved fit for Proxima, although with larger errors for some of the astrometric parameters.\looseness=-4

We report in Table~\ref{tab:fitcolcor} the results of the astrometric fit with the chromatic-correction terms in the model. The final $\chi^2_{\rm r}$ is 0.48. The agreement with \gaia is slightly better (0.5-$\sigma$ level). The largest discrepancies are again $\Delta\alpha_{2016} \sim 0.1$ mas and $\Delta\varpi \sim 0.2$ mas. Convergence of the astrometric-fit process was reached in six iterations, and the number of rejected measurements in the process was 9. Using this solution and the same prescriptions detailed in Sect.~\ref{sec:pma}, we obtain a mass for Proxima\,c of $m_c = 2.5^{+5.1}_{-2.5}$ $M_\oplus$.\looseness=-4

\section{Corner plots}\label{appendix:corner}

Figure~\ref{fig:corner} displays the corner plots of the final fit of the astrometric parameters of Proxima described in Sect.~\ref{sec:astromodel}.\looseness=-4

\section{Observed positions of Proxima.}\label{appendix:positions}
Table~\ref{tab:positions} provides the measurements of Proxima included in our fit, together with other useful quantities. We provide both the pixel-based positions corrected for GD in each image (i.e., those actually used in Sect.~\ref{sec:astromodel}) and the corresponding Equatorial coordinates.\looseness=-4

\begin{sidewaystable*}[t!]
\centering
\caption{Proxima's measurements used in our work.\looseness=-4}
\begin{tabular}{ccccccccccccc}
\hline
ID & Epoch & $x_{\rm c}$ & $\sigma_{x_{\rm c}}$ & $y_{\rm c}$ & $\sigma_{y_{\rm c}}$ & $m$ & Filter & \texttt{QFIT} & $\alpha$ & $\sigma_{\alpha}$ & $\delta$ & $\sigma_{\rm delta}$ \\
& {[Julian Date]} & [pixel] & [pixel] & [pixel] & [pixel] & & & & [deg] & [mas] & [deg] & [mas] \\
\hline
01 & 2459804.75467 & 1926.2235 & 0.0338 & 3286.1748 & 0.0352 & $-$13.057 & F467M & 0.022 & 217.376747371 & 1.351 & $-$62.674587667 & 1.407 \\
02 & 2459632.39110 & 2264.7612 & 0.0332 & 1247.9268 & 0.0320 & $-$13.059 & F467M & 0.028 & 217.378668158 & 1.327 & $-$62.674865292 & 1.282 \\
03 & 2459632.47387 & 2486.6214 & 0.0352 & 1438.5702 & 0.0340 & $-$13.067 & F467M & 0.028 & 217.378667999 & 1.409 & $-$62.674865318 & 1.361 \\
04 & 2460024.85196 & 2385.5600 & 0.0314 & 1421.2092 & 0.0319 & $-$13.065 & F467M & 0.021 & 217.376087566 & 1.255 & $-$62.674683086 & 1.277 \\
05 & 2460172.73301 & 1800.7917 & 0.0318 & 3118.5070 & 0.0321 & $-$13.114 & F467M & 0.031 & 217.374444829 & 1.272 & $-$62.674365571 & 1.283 \\
06 & 2460172.76753 & 2022.4289 & 0.0319 & 3309.6327 & 0.0314 & $-$13.115 & F467M & 0.027 & 217.374444218 & 1.275 & $-$62.674365520 & 1.257 \\
07 & 2460358.36622 & 2062.9079 & 0.0326 & 1234.5178 & 0.0345 & $-$13.040 & F467M & 0.019 & 217.374128875 & 1.304 & $-$62.674429629 & 1.378 \\
08 & 2460358.40075 & 2284.9303 & 0.0315 & 1425.4628 & 0.0300 & $-$13.039 & F467M & 0.029 & 217.374128667 & 1.261 & $-$62.674429719 & 1.200 \\
09 & 2460504.43855 & 1947.3915 & 0.0293 & 3111.0133 & 0.0299 & $-$13.093 & F467M & 0.024 & 217.372376013 & 1.171 & $-$62.674260916 & 1.196 \\
10 & 2460504.47009 & 2168.2866 & 0.0308 & 3302.0919 & 0.0316 & $-$13.086 & F467M & 0.033 & 217.372376032 & 1.231 & $-$62.674260733 & 1.263 \\
11 & 2460725.51673 & 1970.8004 & 0.0332 & 1189.9705 & 0.0330 & $-$13.111 & F467M & 0.031 & 217.371822580 & 1.327 & $-$62.674218487 & 1.322 \\
12 & 2460725.54827 & 2191.9636 & 0.0323 & 1380.8853 & 0.0326 & $-$13.123 & F467M & 0.021 & 217.371821662 & 1.291 & $-$62.674218800 & 1.303 \\
13 & 2456948.47274 & 1050.4672 & 0.0271 & 1190.2116 & 0.0269 & $-$13.557 & F555W & 0.056 & 217.394963223 & 1.085 & $-$62.676168291 & 1.075 \\
14 & 2456948.48598 & 949.4063 & 0.0289 & 1089.8855 & 0.0292 & $-$13.536 & F555W & 0.053 & 217.394962603 & 1.157 & $-$62.676168495 & 1.166 \\
15 & 2456948.49706 & 788.7839 & 0.0341 & 1191.0784 & 0.0346 & $-$13.551 & F555W & 0.056 & 217.394963344 & 1.364 & $-$62.676168466 & 1.385 \\
16 & 2456948.50370 & 788.8014 & 0.0335 & 1191.0997 & 0.0342 & $-$13.547 & F555W & 0.054 & 217.394963371 & 1.341 & $-$62.676168476 & 1.369 \\
17 & 2456749.02448 & 1162.0323 & 0.0335 & 1258.4659 & 0.0324 & $-$13.589 & F555W & 0.054 & 217.396538553 & 1.340 & $-$62.676610051 & 1.297 \\
18 & 2456749.03111 & 1161.1490 & 0.0315 & 1057.3359 & 0.0312 & $-$13.556 & F555W & 0.059 & 217.396539275 & 1.261 & $-$62.676610395 & 1.247 \\
19 & 2456749.04218 & 1035.8800 & 0.0264 & 1158.2595 & 0.0256 & $-$13.582 & F555W & 0.066 & 217.396539564 & 1.057 & $-$62.676610018 & 1.023 \\
20 & 2457128.81620 & 971.4732 & 0.0398 & 1203.8484 & 0.0370 & $-$13.614 & F555W& 0.051 & 217.394057801 & 1.594 & $-$62.676392068 & 1.480 \\
21 & 2457128.82282 & 971.1185 & 0.0276 & 1088.1644 & 0.0273 & $-$13.622 & F555W& 0.044 & 217.394057503 & 1.105 & $-$62.676392077 & 1.094 \\
22 & 2457128.82890 & 1081.6568 & 0.0278 & 1052.5601 & 0.0289 & $-$13.582 & F555W & 0.061 & 217.394057900 & 1.113 & $-$62.676392205 & 1.155 \\
23 & 2457238.36896 & 1031.7854 & 0.0311 & 1154.1121 & 0.0307 & $-$13.581 & F555W & 0.053 & 217.392817136 & 1.244 & $-$62.676113427 & 1.226 \\
24 & 2457238.37558 & 1132.2106 & 0.0344 & 1153.7219 & 0.0350 & $-$13.615 & F555W & 0.049 & 217.392816996 & 1.374 & $-$62.676113579 & 1.400 \\
25 & 2457238.38220 & 1062.2426 & 0.0210 & 1249.5657 & 0.0208 & $-$13.591 & F555W & 0.059 & 217.392816489 & 0.841 & $-$62.676113617 & 0.831 \\
26 & 2457238.38883 & 951.6326 & 0.0287 & 1212.2432 & 0.0291 & $-$13.568 & F555W & 0.059 & 217.392816007 & 1.147 & $-$62.676113547 & 1.164 \\
27 & 2457238.39545 & 951.1152 & 0.0316 & 1096.5853 & 0.0323 & $-$13.567 & F555W & 0.055 & 217.392815855 & 1.264 & $-$62.676113501 & 1.290 \\
28 & 2457238.40153 & 1061.5841 & 0.0341 & 1061.0509 & 0.0326 & $-$13.561 & F555W & 0.052 & 217.392817084 & 1.365 & $-$62.676113205 & 1.303 \\
29 & 2458120.16967 & 950.3832 & 0.0291 & 968.7748 & 0.0293 & $-$13.580 & F555W & 0.049 & 217.388121548 & 1.163 & $-$62.675610140 & 1.173 \\
30 & 2458120.17629 & 1050.0150 & 0.0291 & 968.5612 & 0.0282 & $-$13.579 & F555W & 0.053 & 217.388121423 & 1.162 & $-$62.675610305 & 1.129 \\
31 & 2458120.18291 & 981.1098 & 0.0295 & 1063.1812 & 0.0300 & $-$13.576 & F555W & 0.054 & 217.388121310 & 1.178 & $-$62.675610317 & 1.198 \\
32 & 2458120.18953 & 870.2524 & 0.0290 & 1027.2180 & 0.0308 & $-$13.567 & F555W & 0.065 & 217.388121609 & 1.162 & $-$62.675610342 & 1.231 \\
33 & 2458120.19616 & 869.8791 & 0.0306 & 911.4132 & 0.0304 & $-$13.573 & F555W & 0.063 & 217.388121165 & 1.223 & $-$62.675610106 & 1.217 \\
34 & 2458120.20278 & 980.3003 & 0.0315 & 873.1229 & 0.0307 & $-$13.566 & F555W & 0.064 & 217.388121330 & 1.259 & $-$62.675610364 & 1.227 \\
35 & 2457439.17114 & 1090.4346 & 0.0324 & 1156.1086 & 0.0309 & $-$13.535 & F555W & 0.058 & 217.392412415 & 1.297 & $-$62.676143506 & 1.235 \\
36 & 2457439.19546 & 829.0430 & 0.0290 & 1157.1928 & 0.0291 & $-$13.531 & F555W & 0.070 & 217.392412132 & 1.162 & $-$62.676144078 & 1.162 \\
37 & 2457439.20211 & 829.0055 & 0.0289 & 1157.3093 & 0.0291 & $-$13.554 & F555W & 0.064 & 217.392412291 & 1.156 & $-$62.676144024 & 1.163 \\
38 & 2457546.97462 & 1032.8918 & 0.0302 & 1135.9872 & 0.0301 & $-$13.543 & F555W & 0.048 & 217.391061298 & 1.209 & $-$62.676082565 & 1.203 \\
39 & 2457546.98570 & 872.2630 & 0.0291 & 1237.2087 & 0.0286 & $-$13.686 & F555W & 0.062 & 217.391061082 & 1.162 & $-$62.676082535 & 1.145 \\
40 & 2457651.62777 & 1138.9173 & 0.0307 & 1185.5691 & 0.0308 & $-$13.560 & F555W & 0.062 & 217.390377397 & 1.228 & $-$62.675772426 & 1.233 \\
41 & 2457651.64101 & 1169.7973 & 0.0295 & 1279.9915 & 0.0289 & $-$13.575 & F555W & 0.062 & 217.390376915 & 1.181 & $-$62.675772452 & 1.156 \\
42 & 2457651.64763 & 1058.8164 & 0.0305 & 1243.8295 & 0.0312 & $-$13.551 & F555W & 0.063 & 217.390376895 & 1.219 & $-$62.675772244 & 1.247 \\
43 & 2457651.65425 & 1058.3481 & 0.0296 & 1127.9507 & 0.0306 & $-$13.537 & F555W & 0.065 & 217.390376776 & 1.185 & $-$62.675772600 & 1.224 \\
44 & 2457754.71986 & 909.7572 & 0.0290 & 1000.7889 & 0.0291 & $-$13.552 & F555W & 0.042 & 217.390409935 & 1.159 & $-$62.675823525 & 1.164 \\
45 & 2458014.98180 & 1231.9042 & 0.0312 & 1198.2049 & 0.0303 & $-$13.543 & F555W & 0.049 & 217.388091362 & 1.250 & $-$62.675562392 & 1.214 \\
46 & 2458015.00166 & 1151.7625 & 0.0333 & 1256.6533 & 0.0318 & $-$13.533 & F555W & 0.056 & 217.388090579 & 1.331 & $-$62.675562181 & 1.270 \\
47 & 2458015.01490 & 1261.7607 & 0.0323 & 1102.7590 & 0.0322 & $-$13.563 & F555W & 0.050 & 217.388090559 & 1.290 & $-$62.675562552 & 1.287 \\
\hline
\end{tabular}
\tablefoot{ID is an arbitrary row index. Positions and corresponding errors are in WFC3/UVIS pixels. $m$ is the instrumental magnitude ($-2.5\log_{10}{\rm Flux}$). \texttt{QFIT} is the quality of the PSF fit.}
\label{tab:positions}
\end{sidewaystable*}

\end{appendix}

\end{document}